\documentclass[aps,prd,twocolumn,nofootinbib,showpacs,superscriptaddress]{revtex4-1}
\usepackage{amsmath,amsfonts,amssymb,bm}
\usepackage{graphicx}
\usepackage{subfigure}
\usepackage{color}
\usepackage{multirow} %for tabular

\usepackage{epstopdf}

\definecolor{purple}{rgb}{0.5,0,0.5}
\definecolor{blue}{rgb}{0.0,0,0.9}
\usepackage[colorlinks=true, pdfstartview=FitV, linkcolor=purple, citecolor= purple, urlcolor=blue]{hyperref}

\begin{document}

\title{Chemical Freeze-out Parameters via a Non-perturbative QCD Approach}

\author{Yi Lu}
\email{qwertylou@pku.edu.cn}
\affiliation{Department of Physics and State
Key Laboratory of Nuclear Physics and Technology, Peking
University, Beijing 100871, China}
\affiliation{Collaborative Innovation Center of Quantum Matter, Beijing 100871, China}

\author{Muyang Chen }
%\affiliation{Department of Physics and State
%Key Laboratory of Nuclear Physics and Technology, Peking University, Beijing 100871, China}
\affiliation{Department of Physics, Hunan Normal University, Changsha 410081, China}

\author{Zhan Bai}
\email[Present address: ]{{Institute of Theoretical Physics, Chinese }\\  {Academy of Science, Beijing 100081, China}}
\affiliation{Department of Physics and State
Key Laboratory of Nuclear Physics and Technology, Peking University, Beijing 100871, China}
\affiliation{Collaborative Innovation Center of Quantum Matter, Beijing 100871, China}

\author{Fei Gao }
\affiliation{Institut f{\"u}r Theoretische Physik, 	Universit{\"a}t Heidelberg, Philosophenweg 16,
	69120 Heidelberg, Germany }

\author{Yu-xin Liu }
\email[Corresponding author: ]{yxliu@pku.edu.cn}
\affiliation{Department of Physics and State Key Laboratory of
Nuclear Physics and Technology, Peking University, Beijing 100871,
China}
\affiliation{Collaborative Innovation Center of Quantum Matter, Beijing 100871, China}
\affiliation{Center for High Energy Physics, Peking
University, Beijing 100871, China}

\begin{abstract}
By analyzing the calculated baryon number susceptibility ratios ${\chi_{1}^{B}}/{\chi_{2}^{B}}$ and ${\chi_{3}^{B}}/{\chi_{1}^{B}}$ in two-flavor system via the Dyson-Schwinger equation approach of QCD, we determine the chemical freeze-out temperature and baryon chemical potential in cases of both thermodynamic limit and finite size.
We calculate the center-of-mass energy dependence of the ${\chi_{4}^{B}}/{\chi_{2}^{B}}\, (\kappa \sigma^{2})$ at  the freeze-out line and find an excellent agreement with experimental data
%in $\sqrt{S_{NN}^{}} \geq 19.6\,$GeV region
when taking into account the finite size effect.
Our calculations indicate that the $\kappa \sigma^{2}$ exhibits a nonmonotonic behavior in lower collision energy region.
We also predict that the collision energy dependence of ${\chi_{6}^{B}}/{\chi_{2}^{B}}$ is nonmonotonic.
\end{abstract}

%\pacs{25.75.Nq, 11.10.Wx, 12.38.Lg, 21.65.Qr }

\maketitle

%%%%%
\section{introduction}\label{sec:Intro}

Phase transitions of strong interaction matter have been explored for more than forty years since the research may reveal the nature of the early universe matter evolution~\cite{BraunMunzinger:2009zz,Philipsen:2012nu,Gupta:2011wh}.
The transitions include chiral phase transition (from dynamical chiral symmetry to dynamical chiral symmetry breaking) which generates more than $98\%$ of the mass of visible matter
and the confinement transition (hadronization) which slaves the quarks and gluons to hadrons.
They are driven by the temperature ($T$) and the baryon density ($\rho_{B}^{}$) or chemical potential ($\mu_{B}^{}$).
Since the strong interaction can be well described by Quantum Chromodynamics (QCD),
the above mentioned phase transitions are usually referred to as QCD phase transitions.
Moreover, many calculations (see,  {\it e.g.},  Refs.~\cite{Aoki:2006we,Aoki:2009sc,Ejiri:2008xt,Gupta:2011wh,Philipsen:2012nu,Li:2011ee,Qin:2010nq,Xin:2014ela,Fischer:2011mz,Fischer:2012vc,Fischer:2014ata,Gao:2015kea,Ratti:2005jh,Schaefer:2007pw,Fu:2007xc,Fukushima:2008wg,Jiang:2013yoa,Xin:2014dia}) have shown that the chiral phase transition at low chemical potential is a crossover at physical quark mass.
Theoretical calculations (see,  {\it e.g.},  Refs.~\cite{Ratti:2005jh,Schaefer:2007pw,Fu:2007xc,Ejiri:2008xt,Gupta:2011wh,Fukushima:2008wg,Qin:2010nq,Xin:2014ela,Xin:2014dia,Fischer:2011mz,Fischer:2012vc,Fischer:2014ata,Gao:2015kea,Jiang:2013yoa,Xin:2014dia}) also indicate that the chiral phase transition at high chemical potential is first order.
Therefore, there would exist a critical end-point (CEP) in the $T$--$\mu_{B}^{}$ plane
at which the first order phase transition turns to crossover.
The position of the CEP or even its existence becomes thus one of the most significant topic in both theories and experiments.
Besides the efforts in theories,
the Beam Energy Scan (BES) program at RHIC, the FAIR at GSI and the NICA at DUBNA  all take the search of the CEP as their investigation focus (see, {\it e.g.}, Refs.~\cite{Melkumov:2011zz,Melkumov:2012zz,Odyniec:2012zza}) and some meaningful information has been provided by the RHIC experiments~\cite{Adamczyk:2013dal,Adamczyk:2014fia,Luo:2015ewa,STAR:2021kur}.

In experiments, one can measure only the states after the hadronization but not the phase transition directly, and thus the chemical freeze-out line which is defined as the set of states ceasing the inelastic collision of the newly formed hadrons plays the essential role.
Especially, as the chemical freeze-out line  approaches to the CEP, nonmonotonic behavior of conserved charge fluctuations could be observed~\cite{Stephanov:1998dy,Stephanov:2008qz,Stephanov:2011pb, Stephanov:1999zu,Stephanov:2011zz,Hatta:2003wn,Xin:2014ela}.
The freeze--out temperature and chemical potential have then been studied in statistical hadronization model (SHM)~\cite{Becattini:2005xt,Andronic:2005yp,Andronic:2008gu,Das:2014kja,Das:2014oca}, hadron resonance gas (HRG) model~\cite{Karsch:2010ck,Alba:2014eba},
lattice QCD simulations~\cite{Gavai:2010zn,Bazavov:2012vg,Borsanyi:2013hza,Borsanyi:2014ewa} and other models~\cite{Cleymans:2005xv,Chen:2015dra}.
In fact, the matter system generated in relativistic heavy ion collision (RHIC) experiment has a finite size and cools in a finite time~\cite{Stephanov:1999zu,Stephanov:2011zz,Berdnikov:1999ph,Abelev:2014pja}.
The finite size and finite time prevent the correlation length $\xi$ from diverging near the CEP, and smoothen the fluctuations~\cite{Berdnikov:1999ph}.
Model calculations have shown that the finite size influences both the phase diagram and the thermodynamical properties  drastically~\cite{Palhares:2009tf,Magdy:2015eda,Shao:2006gz,Bhattacharyya:2012rp,Bhattacharyya:2014uxa,Bhattacharyya:2015kda},  the surface of the system may also play the role~\cite{Berger:1986ps,Deutsch:1978sc,Elze:1986db,Ke:2013wga,Gao:2016interface,Zhao:2019fs}.
The effects of the finite size and the surface on the chemical freeze-out parameters  will then complement the information for searching the CEP in experiments.
However, different models give contradictory results.
It is therefore imperative to investigate the finite size and the surface effects on the chemical freeze-out parameters with sophisticated QCD  approaches.

It has been known that Dyson-Schwinger equations (DSE), a nonperturbative approach of QCD
~\cite{Roberts:1994dr,Roberts:2000aa,Maris:2003vk,Alkofer:2000wg,Bashir:2012fs,Cloet:2013jya,Chang:2010hb,Qin:2013mta,Qin:2011dd},
have been successful in describing  QCD phase transitions (see, {\it e.g.}, Refs.~\cite{Bashir:2012fs,Qin:2010nq,Xin:2014ela,Qin:2010pc,Fischer:2009wc,Fischer:2009gk,Fischer:2011mz,Fischer:2012vc,Fischer:2014ata,Gao:2015kea,Gutierrez:2013sta,Wang:2014yla,Wang:2013wk}) and hadron properties (For recent reviews, see Refs.~\cite{Bashir:2012fs,Cloet:2013jya}).
We then, in this paper, take the DSE approach to investigate the chemical freeze-out parameters with  the finite size and surface effects being taken into account. We calculate the baryon number susceptibilities in two light-flavor quark system. By comparing the obtained  baryon number susceptibility ratios ${\chi_{1}^{B}}/{\chi_{2}^{B}}$ and ${\chi_{3}^{B}}/{\chi_{1}^{B}}$ with the experimental data of the net-proton distribution cumulant ratios $C_1/C_2$ and $C_3/C_1$ at different collision energies, we determine the freeze-out parameters.
We observe that with the finite size and surface effects being included, the calculated collision energy dependence of the ${\chi_{4}^{B}}/{\chi_{2}^{B}}$ agrees with the experimental data excellently,
%in $\sqrt{S_{NN}^{}} \geq 19.6\,$GeV region excellently.
and the calculated $\kappa \sigma^2$ shows a nonmonotonic behavior in lower collision energy region.
Moreover, we propose that hyper-order cumulant ratio such as ${\chi_{6}^{B}}/{\chi_{2}^{B}}$ also shows a nonmonotonic dependence on the collision energy.
%
%value of ${\chi_{4}^{B}}/{\chi_{2}^{B}}$ at $\sqrt{S_{NN}} = 5.5\;$GeV.
%

The remainder of this paper is organized as follows:
In Sec.~\ref{sec:Theory}, we describe briefly the Dyson-Schwinger equation approach and its relation to the chemical freeze-out parameters.
In Sec.~\ref{sec:FreezeOutParameters}, we calculate the freeze-out parameters by adopting the DSE approach as well as the experimental data.
In Sec.~\ref{sec:PhaseDiagram}, we give the phase diagram in the vicinity of CEP and reveal the effect of finite size and surface.
In Sec.~\ref{sec:Summary}, we give a summary and discussion.

\section{Theoretical Framework.}\label{sec:Theory}
\subsection{Dyson-Schwinger Equation Approach}\label{sec:DSE}

The Dyson-Schwinger equations are inifinite number of coupled equations.
In this paper, we focus on the DSE for quark propagator $S(\tilde{\omega}_{j},\vec{p})$.
The corresponding equation is:
\begin{equation}\label{eq:gapeq}
S^{-1}(\tilde{\omega}_{j}^{}, \vec{p}) = Z_{2}\left(i \vec{\gamma}\cdot\vec{p} \!+i {\gamma_{4}^{}} \tilde{\omega}_{j}^{}\right) \!+Z_4 m_{0}^{}\! +Z_1\Sigma\left(\tilde{\omega}_j,\vec{p}\right),
\end{equation}
where $Z_{1}$, $Z_{2}$ and $Z_{4}$ are renormalization constants.
$m_{0}$ is the current quark mass.
$\tilde{\omega}_{j}=\omega_{j}+i\mu_q$, with $\mu_{q}$ being the quark chemical potential,
and $\omega_j=(2j+1)\pi T$ the Matsubara frequency for quarks.
$\Sigma(\tilde{\omega}_j,\vec{p})$ is the self-energy of quark, and reads:
\begin{equation}
\begin{split}
\Sigma\left(\tilde{\omega}_j,\vec{p}\right)=&\frac{4}{3}T {\sum\limits_{l=-\infty}^{\infty}}\int\frac{d^3\vec{q}}{(2\pi)^3} g^2 D_{\mu\nu}^{}(\vec{k},\Omega_{jl}^{};T,\mu_{q}^{}) \\
&\times\gamma_{\mu}^{} S(\tilde{\omega}_{l}^{},\vec{q}) \Gamma_{\nu}^{}(\vec{p},\tilde{\omega}_{j}^{},\vec{q},\tilde{\omega}_{l}^{};T,\mu_{q}^{}) \, ,
\end{split}
\end{equation}
where $D_{\mu\nu}^{}$ is the dressed-gluon propagator,
$\Gamma_{\nu}^{}$ is the dressed quark-gluon vertex,
$\Omega_{jl}^{} = \omega_{j}^{} - \omega_{l}^{}$ is the Matsubara frequency for gluon.

In principle, the gluon propagator $D_{\mu\nu}$ and the quark-gluon vertex $\Gamma_{\nu}$ should be solved by corresponding DSEs,
which depend on higher order correlation functions. And a truncation must be applied in order for numerical solution.
In this paper, for the quark-gluon vertex, we adopt at first stage the rainbow approximation for the vertex $\Gamma_{\nu}^{}(\vec{p},\tilde{\omega}_{m}^{},\vec{q},\tilde{\omega}_{l}^{};T,\mu_{q}^{} ) = \gamma_{\nu}^{}$.

The gluon propagator has the general form
\begin{equation}
     g^2D_{\mu\nu}(\Omega_{nl},\vec{k}) = P_{\mu\nu}^{T} D_{T}(\Omega_{nl}^2,\vec{k}^2) + P_{\mu\nu}^{L} D_{L}(\Omega_{nl}^2,\vec{k}^2),
\end{equation}
where $P_{\mu\nu}^{T,L}$ are the transverse and longitudinal projection operators, respectively:
\begin{equation}
\begin{split}
    P_{\mu\nu}^{T} &= (1-\delta_{\mu 4})(1-\delta_{\nu 4})\left(\delta_{\mu\nu}-\frac{k_\mu k_\nu}{k^2}\right) ,\\
    P_{\mu\nu}^{L} &= \left(\delta_{\mu\nu}-\frac{k_\mu k_\nu}{k^2}\right) - P_{\mu\nu}^{T}\, ,
\end{split}
\end{equation}
where $k=\left(\Omega_{nl},\vec{k}\right)$.
$D_{T}$ and $D_{L}$ are the effective interactions and can be represented using models.
Note that the coupling constant $g$ and the renormalization constant $Z_{1}$ has been absorbed into the effective interaction.

In this paper, we adopt the infrared constant model (QC model)~\cite{Qin:2011dd,Xin:2014ela,Gao:2015kea}, which reads:
\begin{equation}
     \begin{split}
     \mathcal{D}(s)=&8\pi^2\frac{D}{\omega^4}\textrm{e}^{-s/\omega^2}+\alpha_{\textrm{pQCD}}(s),
     \end{split}
\end{equation}
where $D$ and $\omega$ are the parameters of the model.
$\alpha_{\textrm{pQCD}}$ is the ultraviolet perturbation term and reads:
\begin{equation}
     \alpha_{\textrm{pQCD}}(s) = \frac{8\pi^2\gamma_{m}}{\ln\left[\tau+\left(1+s/\Lambda^2_{\textrm{QCD}}\right)^2\right]}\mathcal{F}(s),
\end{equation}
where $\mathcal{F}(s)=\left[1-\exp\left(-s/4m_t^2\right)\right]$,
$\tau=e^2-1$, $m_t=0.5\;$GeV, $\Lambda_{\textrm{QCD}}=0.234\;$GeV, and $\gamma_{m}=12/(33-N_f)$ with $N_f=4$.

The gluon screening mass $m_g$ is also considered in the longitudinal part of gluon model~\cite{Gao:2015kea,Gao:2016susc}:
\begin{equation}
\begin{split}
    D_{T}(\Omega_{nl}^2,\vec{k}^2) &= \mathcal{D}(\Omega_{nl}^2 + \vec{k}^2) ,\\
    D_{L}(\Omega_{nl}^2,\vec{k}^2) &= \mathcal{D}(\Omega_{nl}^2 + \vec{k}^2 + m_g^2),
\end{split}
\end{equation}
whose value is determined by leading-order perturbative QCD~\cite{Haque:2013gth}:
\begin{equation}
    m_{g}^{2} = \frac{16}{5}\left(T^2+ \frac{6}{5\pi^2}\mu_q^2 \right).
\end{equation}

In order to fix the renormalization constant $Z_2$ and $Z_4$,
we need to specify the renormalization condition. In this paper, the renormalization condition is:
\begin{equation}
S^{-1}(p^2)\bigg|_{\omega_0^2+\vec{p}^2=\zeta^2}=i\left(\vec{\gamma}\cdot\vec{p}+\gamma_4\tilde{\omega}_0\right)+m_0,
\end{equation}
where $\zeta$ is the renormalization point. We choose $\zeta=19\;$GeV, $m_{0}^{}\!=\!3.4\,$MeV, $D=1.024\,\textrm{GeV}^2$ and $\omega\!=\!0.5\,$GeV as in Ref.~\cite{Qin:2011dd}.

The quark propagator can be decomposed according to its Lorentz structure.
At finite temperature, the decomposition is:
\begin{equation}\label{eq:LorentzDecomposition}
     \begin{split}
     S(\tilde{\omega}_j,\vec{p})^{-1}=&i\vec{\gamma}\cdot\vec{p}A(\tilde{\omega}_j^2,\vec{p}^{\,2})\\
     &+i\gamma_4\tilde{\omega}_jC(\tilde{\omega}_j^2,\vec{p}^{\,2})+B(\tilde{\omega}_j^2,\vec{p}^{\,2}).
     \end{split}
\end{equation}
There should be, in principle, a fourth term in this decomposition.
However, its contribution to order parameter is extremely small and is usually omitted in practical calculations~\cite{Roberts:2000aa,Contant:2017PRD}.

The mass function of the quark propagator can then be defined as:
\begin{equation}\label{eq:MassFunction}
     M(\tilde{\omega}_j^2,\vec{p}^{\,2})=B(\tilde{\omega}_j^2,\vec{p}^{\,2})/A(\tilde{\omega}_j^2,\vec{p}^{\,2}).
\end{equation}
In vacuum, the physical solution to the quark DSE has a non-zero mass function even if the current quark mass is zero.
At high temperature, the mass function gradually approaches the current quark mass.
Therefore, the mass function at zero momentum, $M(\tilde{\omega}_0^2,0)$, is often used as the order parameter of the QCD crossover.

\subsection{Number Density and susceptibilities}
Experimental observations indicate that the yields of pion and proton are much larger than that of kaon~\cite{Das:2014kja}, we can then simplify the matter generated in RHIC experiments as that including mainly two light flavor quarks. In the system of $u$ and $d$ quarks, baryon number density $n_{B}^{}$ and electric charge density $n_{Q}^{}$ can be fixed with quark number density $n_{u,d}^{}$ as:
\begin{equation}\label{eq:baryonnumber}
 n_{B}^{} = \frac{1}{3}n_{u}^{} + \frac{1}{3}n_{d}^{} \, , \qquad
 n_{Q}^{} = \frac{2}{3}n_{u}^{} - \frac{1}{3}n_{d}^{} \, .
\end{equation}
From Eq.~(\ref{eq:baryonnumber}) we notice that the $u$ and $d$ quarks are in exact isospin symmetry, if only the baryon number is considered.
In this sense, both the $u$ quark and $d$ quark hold the same quark chemical potential $\mu_{q}^{} =\mu_{B}^{}/3$, and the quark number density $n_{q}^{} = 3n_{B}^{}$.

In view of statistical physics, the quark number density can be determined as
\begin{equation}\label{eq:qnumberdistribution}
 n_{q}^{}(\mu_q,T)  =  2 N_{c} N_{f} Z_{2} \! \int \! \frac{d^3 \vec{p}}{(2\pi)^3} f_1(|\vec{p}|; \mu_{q}^{}, T)\, , \quad
\end{equation}
\begin{equation}\label{eq:fermiondistribution}
 f_{1}^{}(|\vec{p}|; \mu_{q}^{}, T)  =  \frac{T}{2} \! \sum_{j=-\infty}^\infty \!\! \textrm{tr}_{D}^{}\left[-\gamma_{4}^{} S(\tilde{\omega}_{j}^{}, \vec{p})\right],
\end{equation}
where $Z_{2}$ is the quark wave-function renormalization constant, $N_{c} = 3$ the color number,
and $N_{f} = 2$ the flavor number.
Notice that the flavor number here represents the flavor degeneracy, and is different from the one we used in the ultraviolet perturbation term in the gluon model.

In Eq.~(\ref{eq:fermiondistribution}), the summation runs over an infinite number of Matsubara frequencies.
In practice, we can only carry out the calculation with a finite number of Matsubara frequency, and the summation must have a cut-off, $N$.
However, the convergence of summation Eq.~(\ref{eq:fermiondistribution}) is slow, especially at low temperature.
%
%In order to solve this problem, we notice that at large momentum, the quark propagator approaches free propagator,
%{\it i.e.}, $A=C=1$, $B=m_0$, where $A$, $B$ and $C$ are the scalar functions defined in Eq.~(\ref{eq:LorentzDecomposition}).
%
On the other hand, we observe that the scalar functions $A$, $B$ and $C$ defined in Eq.~(\ref{eq:LorentzDecomposition}) converge quickly to the free-quark-propagator scalar functions $A=C=1$, $B=m_0$, as the Matsubara frequency grows large.
For free propagator, the distribution function is:
\begin{equation}
     \begin{split}
f^{\textrm{free}}(|\vec{p}|; \mu_{q}^{}, T)&=\,\frac{T}{2}\sum_{j=-\infty}^{\infty}\textrm{tr}_{D}\left[-\gamma_4S^{\textrm{free}}(\tilde{\omega}_j,\vec{p})\right]\\
&=\frac{1}{\textrm{e}^{(E-\mu_q)/T}+1}-\frac{1}{\textrm{e}^{(E+\mu_q)/T}+1},
     \end{split}
\end{equation}
where $E=\sqrt{\vec{p}^{\,2}+m_0^2}$.

Therefore, the contribution of missing Matsubara frequencies that exceed the cut-off $N$ in Eq.~(\ref{eq:fermiondistribution}), can be approximated using the distribution function of free propagator:
\begin{equation}\label{eq:nqtech}
     \begin{split}
f_1 =&\, \frac{T}{2}\sum_{j=-N}^{N}\textrm{tr}_{D}\left[-\gamma_4S(\tilde{\omega}_j,\vec{p})\right]\\
&+f^{\textrm{free}}-\frac{T}{2}\sum_{j=-N}^{N}\textrm{tr}_{D}\left[-\gamma_4S^{\textrm{free}}(\tilde{\omega}_j,\vec{p})\right].
     \end{split}
\end{equation}

\begin{figure}[htb]
    \includegraphics[width=0.45\textwidth]{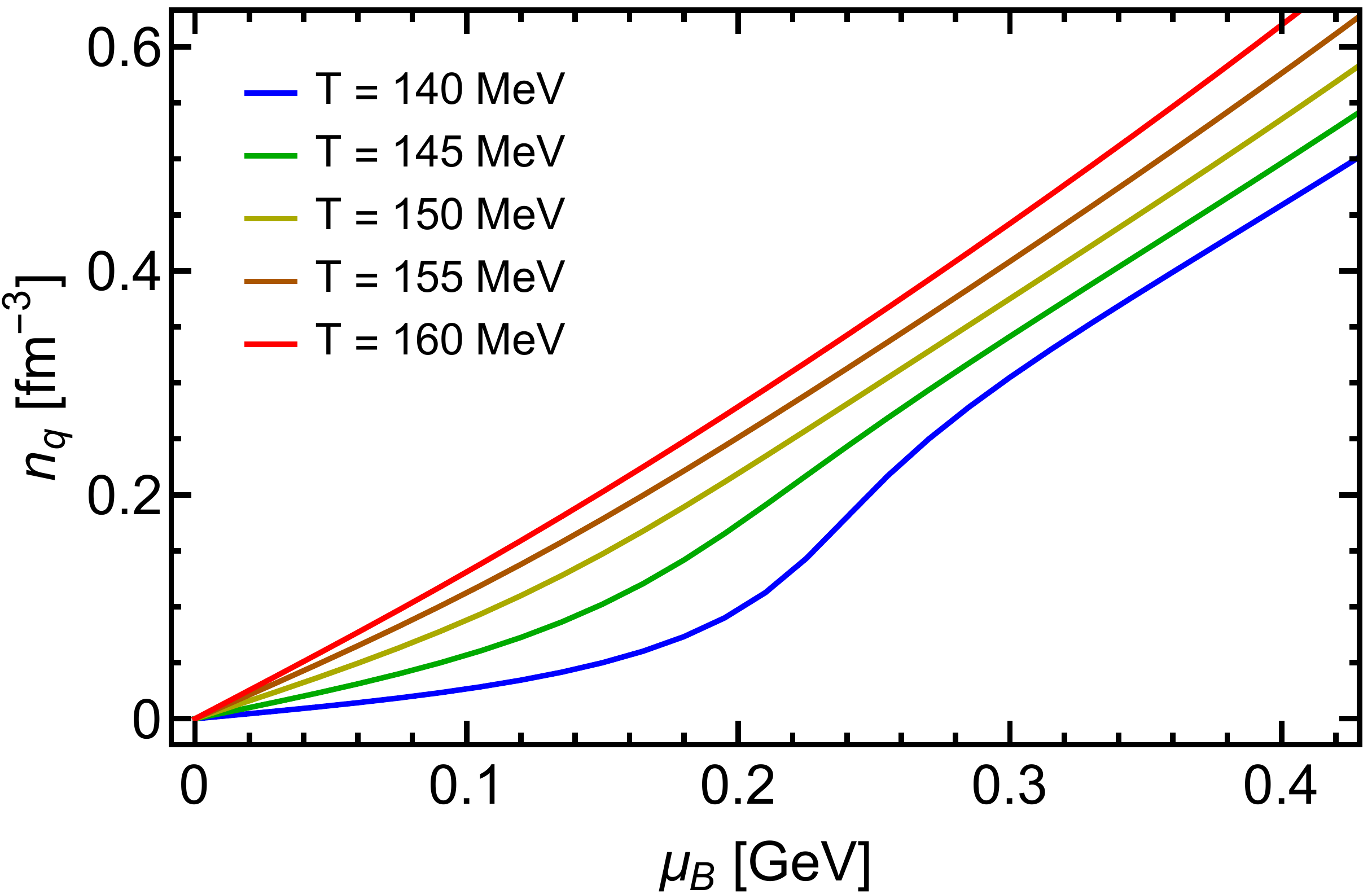}
\caption{\label{fig:nqmu}
Calculated baryon chemical potential dependence of the quark number density at some values of given temperature, in case of thermodynamical limit using Eq.~(\ref{eq:nqtech}).
}
\end{figure}

Using this technique, the calculated quark number density has better convergence at moderate and high temperature, whose chemical potential dependence is shown in Fig.~\ref{fig:nqmu}.

After calculating the quark number density, we can calculate the susceptibility by taking derivatives.
The $k$th order baryon number density susceptibility (fluctuation) is obtained as
\begin{equation}
\begin{split}
 \chi_{1}^{B} &= n_{B} , \\
 \chi_{k}^{B} &= \frac{1}{\beta^{(k-1)}} \frac{\partial^{(k-1)} {n_{B}^{}}}{\partial^{(k-1)} {\mu_{B}^{}}} \, ,
\end{split}
\end{equation}
where $\beta = 1/T$ and $k = 2,3,4,\cdots$.
The susceptibilities are related to the moments of the multiplicity distributions of the corresponding conserved charges as
\begin{eqnarray}
 \nonumber
\frac{\chi_{1}^{}}{\chi_{2}^{}} = M/\sigma^{2} \, ,  \quad &
&\quad \frac{\chi_{3}^{}}{\chi_{1}^{}} = S\sigma^{3}/M \, ,   \\
\frac{ \chi_{3}^{}}{\chi_{2}^{}} = S\sigma \, , \qquad &
&\quad \frac{\chi_{4}^{}}{\chi_{2}^{}} = \kappa \sigma^{2} \, ,
\end{eqnarray}
where $M$, $\sigma^2$, $S$ and $\kappa$ are the mean, the variance, the skewness and the kurtosis of the multiplicity distribution, respectively.
By comparing the theoretical net-baryon number fluctuations in terms of temperature and chemical potential with the experimental data one can  determine the freeze--out parameters~\cite{Borsanyi:2014ewa}.

\subsection{Finite Size and Surface Effect}\label{sec:FiniteSize}

The system created in RHIC exists in finite size, rather than thermodynamical limit.
To determine the freeze--out parameters in experiment one has to take the finite size and the surface effects into account.
Assuming the system as a cube of size $L$, and adopting the anti-periodic condition,
the momentum of a fermion should be $p_{j}^{} = (2j+1)\pi/L$.
%
%However, the momentum quantization depends on the boundary condition of the system.
%In the heavy-ion collision, one should take the spherical boundary condition, and calculation in the framework of discrete momentum is complicated.
%
In principle, one should sum over discrete momentum values.
For simplicity, the finite size effect is roughly incorporated by a non-zero momentum cut-off $|p|_{\textrm{min}}^{} = \pi/L$~\cite{Bhattacharyya:2012rp,Bhattacharyya:2014uxa}.
It corresponds to an infrared momentum cut-off in Eqs.~(\ref{eq:gapeq}) and (\ref{eq:qnumberdistribution}).
This can also be understood by the quantum uncertainty principle, that $L^{-1}$ is the energy scale for a system of finite size $L$.
It is remarkable that such an $L$ is not exactly the same as the size of fireball, but an effective scale that the ingredients of the quark matter can interact.

We also incorporate the effect of the surface through the multiple reflection expansion (MRE) approximation.
In the MRE approximation, the thermodynamical quantities of a droplet composed of quarks can be derived from a density of states in the form~\cite{Berger:1986ps,Deutsch:1978sc,Elze:1986db,Madsen:2000kb,Lugones:2021zya}
\begin{equation}\label{eq:MRE}
 \frac{dN}{dp} = 6\left[ \frac{p^{2} V}{2 \pi^{2}} + f_{s}^{}\left(\frac{p}{M}\right) p S
 + f_{c}^{} \left(\frac{p}{M}\right) C + \cdots \right] \, ,
\end{equation}
where $V$ is the volume of the droplet, $S=4\pi L^2$ and $C=8\pi L$ are the area, the extrinsic curvature of the surface of the droplet, respectively.
The $f_{s}$ and $f_{c}$ are the contributions to the density of states
from the surface and curvature, given explicitly as~\cite{Madsen:2000kb,Shao:2006gz}:
\begin{equation}
     \begin{split}
f_{s}^{} (\frac{p}{M}) &= -\frac{1}{8\pi}\left(1-\frac{2}{\pi}\arctan\left(\frac{p}{M}\right)\right),\\
f_{c}^{} (\frac{p}{M}) &= \frac{1}{12\pi^2}\left[1-\frac{3p}{2M}\left(\frac{\pi}{2}-\arctan\left(\frac{p}{M}\right)\right)\right],
     \end{split}
\end{equation}
with $p$ being the momentum and $M$ the constituent quark mass.

Rigorously, one should solve the coupled equations for the constituent mass $M$~\cite{Shao:2006gz}.
For simplicity we set $M$ in Eq.~(\ref{eq:MRE}) to be $M= \textrm{Re}\,M(\tilde{\omega}_0^2,0)$ from the mass function defined in Eq.~(\ref{eq:MassFunction}), calculated from Eq.~(\ref{eq:gapeq}) with a finite system size $L$.

The modified density of states is then~\cite{Shao:2006gz,Lugones:2021zya}:
\begin{equation}
     \rho_{\textrm{MRE}}^{}(p,M,L)=1+\frac{6\pi^2}{pL}f_{s}+\frac{12\pi^2}{(pL)^2}f_c.
\end{equation}

After taking into account the finite-size as well as the surface effect,
the momentum integration in Sec.~\ref{sec:DSE} should be converted as follows:
\begin{equation}
\int_{0}^{\infty}\frac{\vec{p}^{\,2}\textrm{d}p}{2\pi^2} \longrightarrow \int_{|p|_{\textrm{min}}}^{\Lambda}\rho_{\textrm{MRE}}^{}\frac{\vec{p}^{\, 2}\textrm{d}p}{2\pi^2}.
\end{equation}

%%%%% Results
\section{Freeze-out Parameters}\label{sec:FreezeOutParameters}

\begin{figure}[b]\vspace*{-0.2cm}
    \includegraphics[width=0.45\textwidth]{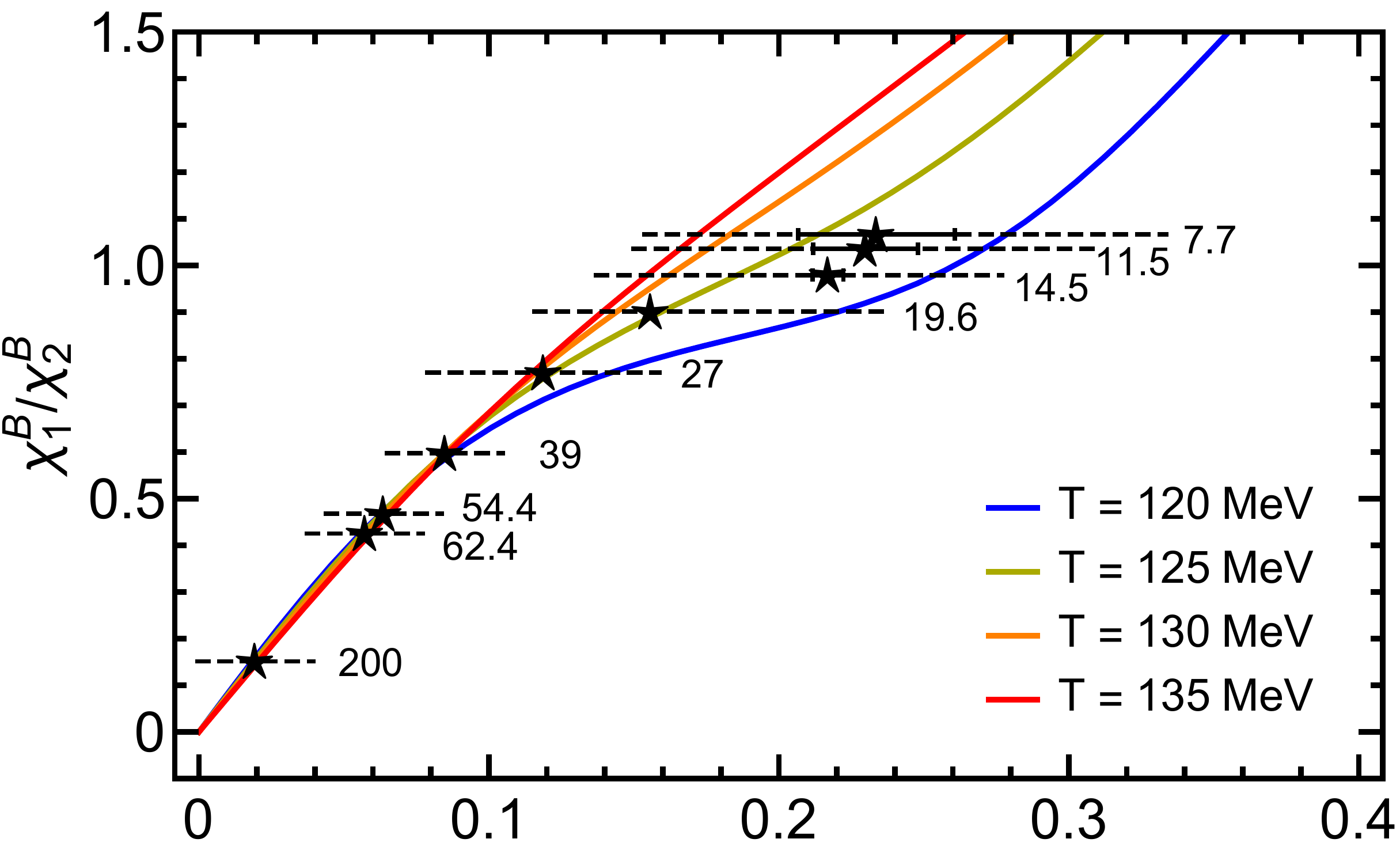}
    \includegraphics[width=0.45\textwidth]{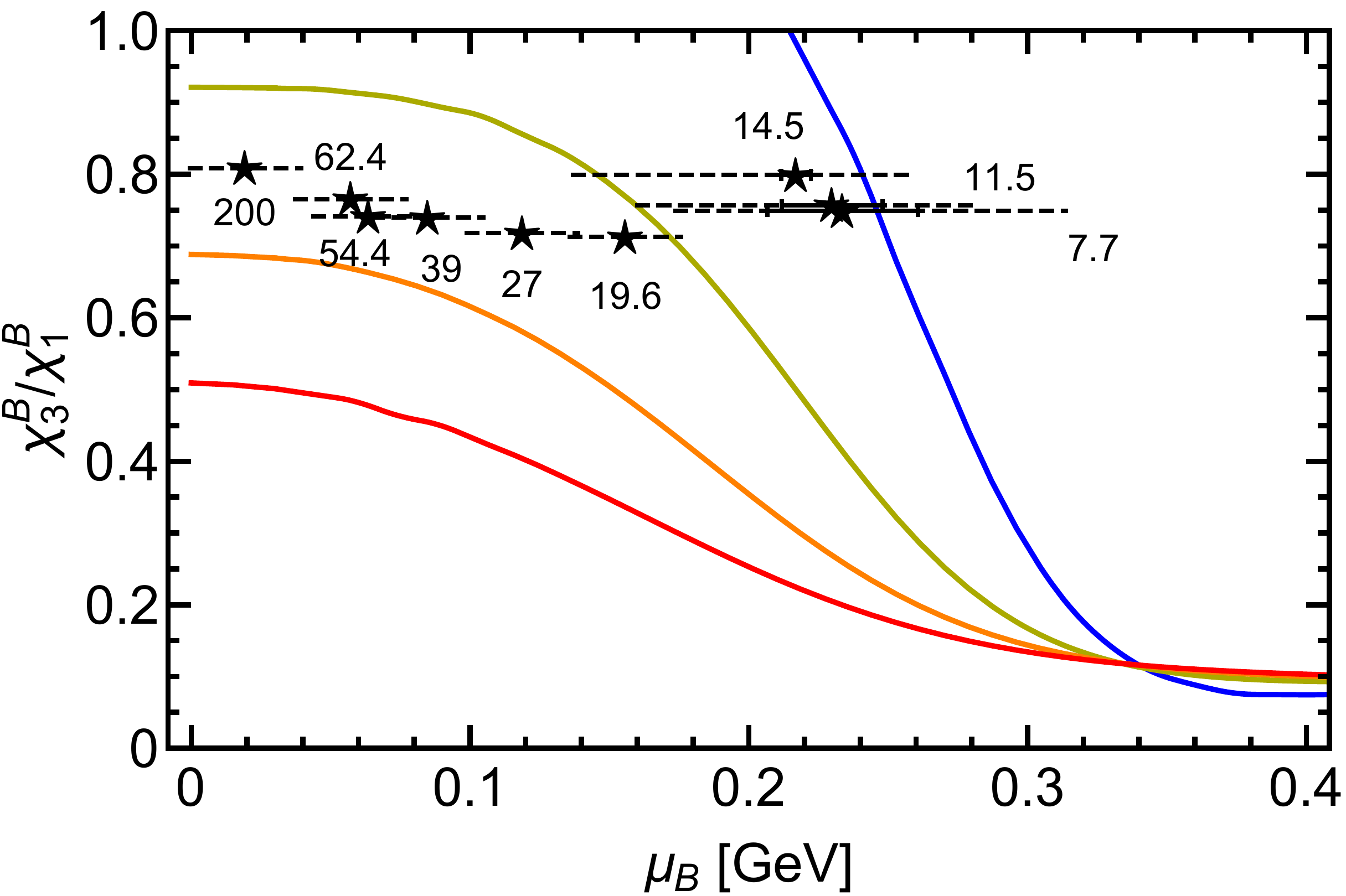}
\caption{\label{fig:x123r2-2} (color online) Calculated baryon chemical potential dependence of the fluctuation ratios  ${\chi_{1}^{B}}/{\chi_{2}^{B}}$ (upper panel)
and ${\chi_{3}^{B}}/{\chi_{1}^{B}}$ (lower panel)
of the system with $L=2.2\,$fm at several values of temperature near the $T_{c}$.
The dashed horizontal lines stand for the experimental values of the efficiency-corrected
$C_{1}/C_{2} = M/\sigma^2$ and $C_{3}/C_{1} = S\sigma^3/M$ of net-proton multiplicity distributions in the central collisions at $\sqrt{S_{NN}^{}} = 200$, $62.4$, $54.4$, $39$, $27$, $19.6$, $14.5$, $11.5$, $7.7\,$GeV given in Ref.~\cite{STAR:2021kur}.
The stars label our assigned freeze-out points.}
\end{figure}

We have carried out calculations with $L=\infty$ (thermodynamical limit) and various finite values of $L$.
The calculations manifest that the fluctuations (skewness, kurtosis, etc.) in the $T$--$\mu_{B}^{}$ plane behave qualitatively the same as those given in Ref.~\cite{Xin:2014ela}, respectively.
%
%,  except for the pseudo-critical temperature $T_{c} $. The $T_{c}^{}$ determined by the chiral susceptibility criterion is $144.0$ MeV in case of $L = \infty$, and $121.6$ MeV %when $L=2.2\,$fm.
%
The obtained $\mu_{B}^{}$ dependence of the baryon number susceptibility ratios  ${\chi_{1}^{B}}/{\chi_{2}^{B}}$
and ${\chi_{3}^{B}}/{\chi_{1}^{B}}$ in case of $L=2.2\,$fm at several values of temperature
are shown in Fig.~\ref{fig:x123r2-2}.
It is evident that our results agree with the lattice QCD results~\cite{Borsanyi:2014ewa} qualitatively very well.
In order to extract the freeze-out parameters, we plot the experimental values of the cumulant ratios $C_{1}/C_{2} = M/ \sigma^2$
and $C_{3}/C_{1} = S\sigma^3/M$ of net-proton multiplicity distributions
in central collisions~\cite{STAR:2021kur} as horizontal lines.

\begin{table*}[t!]
\caption{\label{tab:freezeout} Calculated freeze-out points $( \mu_{B}^{f}, T^{f} )$ in case of different values of $L$ ($T^{f}$ and $\mu_{B}^{f}$ are in unit MeV and $\sqrt{S_{NN}^{}}$ in GeV). }
 \begin{tabular}{|c|c|c|c|c|c|c|c|c|c|c|}
\hline
\multirow{3}{*}{~$\sqrt{S_{NN}}$~} & \multicolumn{2}{|c|}{~~~$L = \textrm{infinity}$~~~} & \multicolumn{2}{|c|}{~~~$L = 3.5\,$fm~~~} & \multicolumn{2}{|c|}{~~~$L = 2.8\,$fm~~~} & \multicolumn{2}{|c|}{~~~$L = 2.5\,$fm~~~} & \multicolumn{2}{|c|}{~~~$L = 2.2\,$fm~~~}\\
%     & \multicolumn{2}{|c|}{($T_{c}=144.0$)} & \multicolumn{2}{|c|}{($T_{c}=???$)} & \multicolumn{2}{|c|}{($T_{c}=???$)} \\
\cline{2-11}
     & $\mu_{B}^{f}$ &   $T^{f}$   & $\mu_{B}^{f}$ &   $T^{f}$   & $\mu_{B}^{f}$ &   $T^{f}$ &$\mu_{B}^{f}$ &   $T^{f}$  &$\mu_{B}^{f}$ &   $T^{f}$   \\
\hline
  200   &   23.7   &  155.5  &   22.8   &   150.0   &   21.9   &   144.1   &  21.1  &  138.8  &  19.4  &  127.5 \\
\hline
  62.4  &   69.9   &  155.1  &   67.1   &   150.0   &   64.4   &   144.0   &  62.1  &  138.9  &  57.2  &  128.0 \\
\hline
  54.4  &   77.2   &  155.1  &   74.6   &   150.0   &   70.2   &   144.3   &  69.1  &  139.1  &  63.8  &  128.4 \\
\hline
  39    &   103.3  &  154.2  &   99.5   &   149.2   &   95.5   &   143.5   &  92.0  &  138.5  &  84.8  &  128.1  \\
\hline
  27    &   154.7  &  150.7  &   143.0  &   147.1   &   135.4  &   141.8   &  129.6 &  137.3  &  118.8 &  127.5 \\
%     &  &  &       &       &       &       \\
\hline
  19.6  &  ~189.1~  &  ~148.0~  &  ~191.7~  &  ~143.1~  &   195.2   &   137.5   &  178.7  &  134.1  &  155.7  & 126.0 \\
        & $\pm$15.7 &  $\pm$1.9 & $\pm$10.9 & $\pm$1.2  & $\pm$2.5  & $\pm$0.2  &         &         &      &  \\
\hline
  14.5  &   ---   &   ---   &   ---   &   ---    &   ~241.5~  &   ~132.0~  &  ~203.3~  &  ~131.6~  &   217.0  &   121.9   \\
        &         &         &         &          & $\pm$21.1  &  $\pm$2.7  & $\pm$23.8 & $\pm$2.2  & $\pm$5.3 & $\pm$0.4  \\
\hline
  11.5  &   ---   &   ---   &   ---   &   ---    &   ---   &   ---   &  ~255.0~  &  ~127.7~  &  ~229.9~  & ~121.4~  \\
        &         &         &         &          &         &         & $\pm$34.3 &  $\pm$2.3 & $\pm$18.0 & $\pm$1.4 \\
\hline
   7.7  &   ---   &   ---   &    ---  &   ---    &   ---   &   ---   &  ~257.4~  &  ~127.5~  &  ~233.7~  & ~121.2~  \\
        &         &         &         &          &         &         & $\pm$39.9 &  $\pm$2.9 & $\pm$27.0 & $\pm$2.1 \\
\hline
\end{tabular}
\end{table*}

By fitting our calculated ${\chi_{1}^{B}}/{\chi_{2}^{B}}$ and ${\chi_{3}^{B}}/{\chi_{1}^{B}}$ values in terms of $T$ and $\mu_{B}^{}$ with the experimental data we get the freeze-out parameters $( \mu_{B}^{f} , T^{f} )$.
The obtained results when $L= \infty$, $3.5\,$fm, $2.8\,$fm, $2.5\,$fm and $2.2\,$fm are listed in Table~\ref{tab:freezeout}.
It appears that our theoretical results in the thermodynamical limit ($L= \infty$) do not fit the experimental values well at low collision energy, whereas the deviations are smaller if the finite size parameter $L$ changes.

\begin{figure}[b]\vspace*{-0.2cm}
    \includegraphics[width=0.45\textwidth]{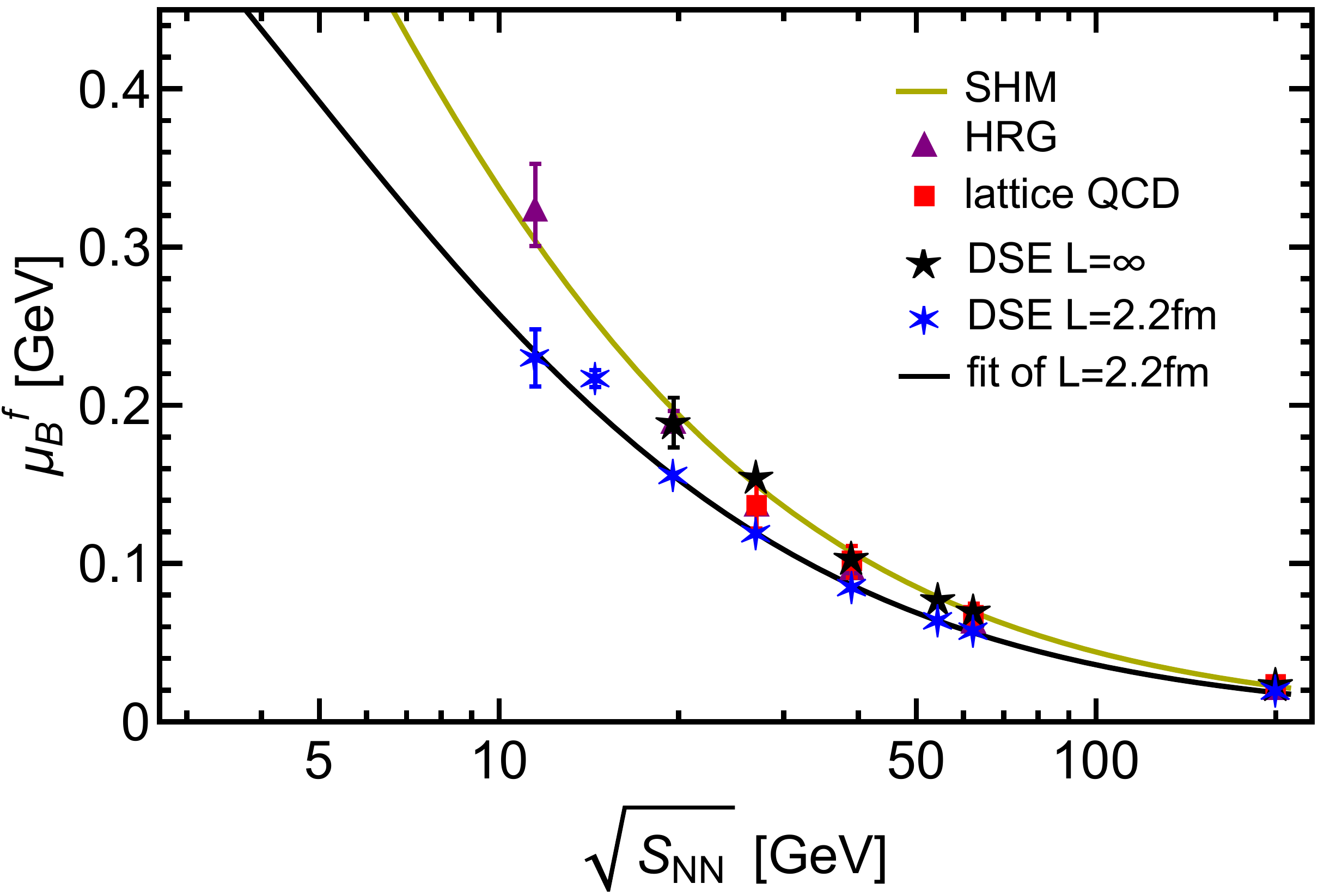}
\caption{\label{fig:mube}
(color online) Comparison of presently obtained $\sqrt{S_{NN}^{}}$ dependence of the baryon chemical potential in cases of $L = \infty$ and $L=2.2\,$fm with those given in lattice QCD simulation~\cite{Borsanyi:2014ewa}, HRG model~\cite{Alba:2014eba} and the parameterized one in SHM model~\cite{Andronic:2005yp}.% The error bars on our fitted line in the $L=2.2\,$fm case label the uncertainty of the freeze-out chemical potential.
}
\end{figure}

We illustrate the presently calculated relation between the baryon chemical potential $\mu_{B}^{f}$ and the center-of-mass energy of the collision,
$\sqrt{S_{NN}^{}}$, and the comparison with those given in lattice QCD simulations ({\it e.g.}, Ref.~\cite{Borsanyi:2014ewa}) and model calculations ({\it e.g.}, Refs.~\cite{Alba:2014eba,Andronic:2005yp}) in Fig.~\ref{fig:mube}.
We see from Fig.~\ref{fig:mube} that our freeze-out baryon chemical potential in case of $L= \infty$  and that when $L=2.2\,$fm match the lattice QCD result and model calculation results well in the region $\mu_{B}^{} < 100\,$MeV,
while those in case of $L=2.2\,$fm deviate from previous results in the $\mu_{B}^{} > 100\,$MeV range.
With the obtained freeze-out points, the freeze-out conditions are fitted as:
\begin{eqnarray}
\label{eq:mapmuBf}
 \mu_{B}^{f} & = & \frac{c}{1+d \sqrt{S_{NN}^{}} } \, ,  \\     \label{eq:mapTf}
 T^{f} &  = & T^{0} \Big[ 1 - a \Big( \frac{\mu_{B}^{f}}{T^{0}} \Big)^2
 - b \Big( \frac{\mu_{B}^{f}}{T^{0}} \Big)^4 \Big] \, .
\end{eqnarray}
Only the freeze-out points with small deviation between theory and experiment in Table~\ref{tab:freezeout} are used for fitting the freeze-out conditions, and the obtained best-fitted parameters are listed in Table~\ref{tab:frzpara}. The fitted $\mu_{B}^{f}(\sqrt{S_{NN}^{}})$ curve is also displayed in Fig.~\ref{fig:mube}.

\begin{table}[t]
\caption{\label{tab:frzpara} Fitted freeze-out parameters $c,d$ in Eq.~\ref{eq:mapmuBf} and $T^0,a,b$ in Eq.~\ref{eq:mapTf}, in case of different finite size parameter $L$ ($c$ and $T^0$ are in unit MeV and $d$ in $\textrm{GeV}^{-1}$).}
 \begin{tabular}{|c|c|c|c|c|c|}
\hline
            ~$L$~             &   ~$c$~   &   ~$d$~   &  ~$T^0$~  &   ~$a$~    &  ~$b$~   \\
\hline
    ~$\textrm{infinity}$~     &   1642.8  &  ~0.373~  &  ~155.7~  &  ~0.0162~  &  ~0.0166~  \\
\hline
         ~$3.5\,$fm~          &   1236.1  &  ~0.285~  &  ~150.3~  &  ~0.0130~  &  ~0.0103~  \\
\hline
         ~$2.8\,$fm~          &   1066.9  &  ~0.256~  &  ~144.6~  &  ~0.0184~  &  ~0.0043~  \\
\hline
         ~$2.5\,$fm~          &  ~840.9~  &  ~0.205~  &  ~139.2~  &  ~0.0137~  &  ~0.0051~  \\
\hline
         ~$2.2\,$fm~          &  ~818.4~  &  ~0.218~  &  ~128.4~  &  ~0.0106~  &  ~0.0022~  \\
\hline
\end{tabular}
\end{table}

With the parametrization, one can predict the freeze-out parameters $({\mu_{B}^{f} }, {T^{f}})$
of the system generated in any collision energy.
For example, with $L=2.2\,$fm, $\sqrt{S_{NN}^{}}=5.8\,$GeV, $7.7\,$GeV, $11.5\,$GeV, $14.5\,$GeV correspond to
$( \mu_{B}^{f}, T^{f} ) = (361.7,99.6)\,\textrm{MeV}$, $(305.8,111.5)\,\textrm{MeV}$, $(233.6,120.8)\,\textrm{MeV}$, $(196.9,123.6)\,\textrm{MeV}$, respectively.

\begin{figure}[b]
    \includegraphics[width=0.45\textwidth]{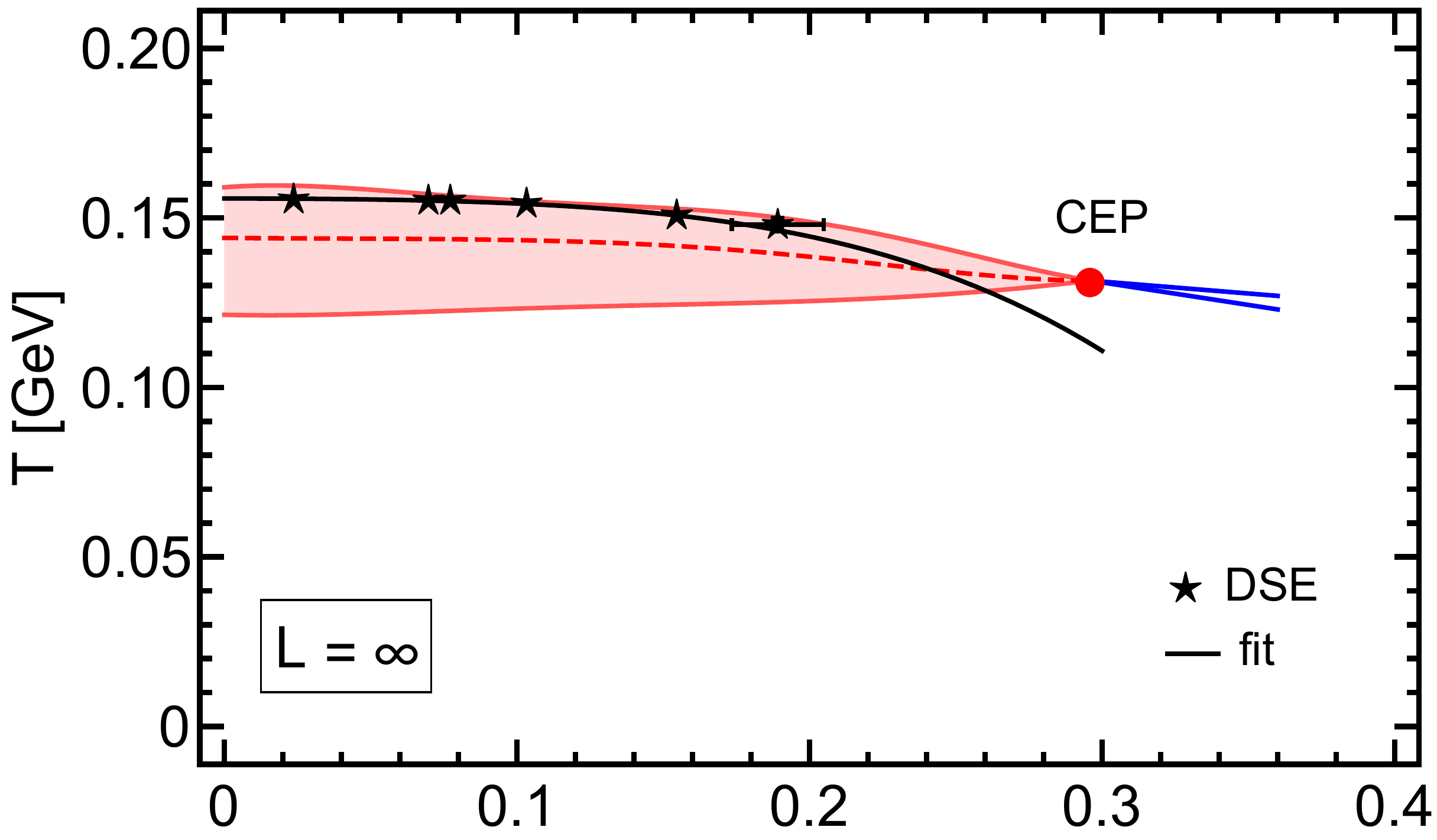}
    \includegraphics[width=0.45\textwidth]{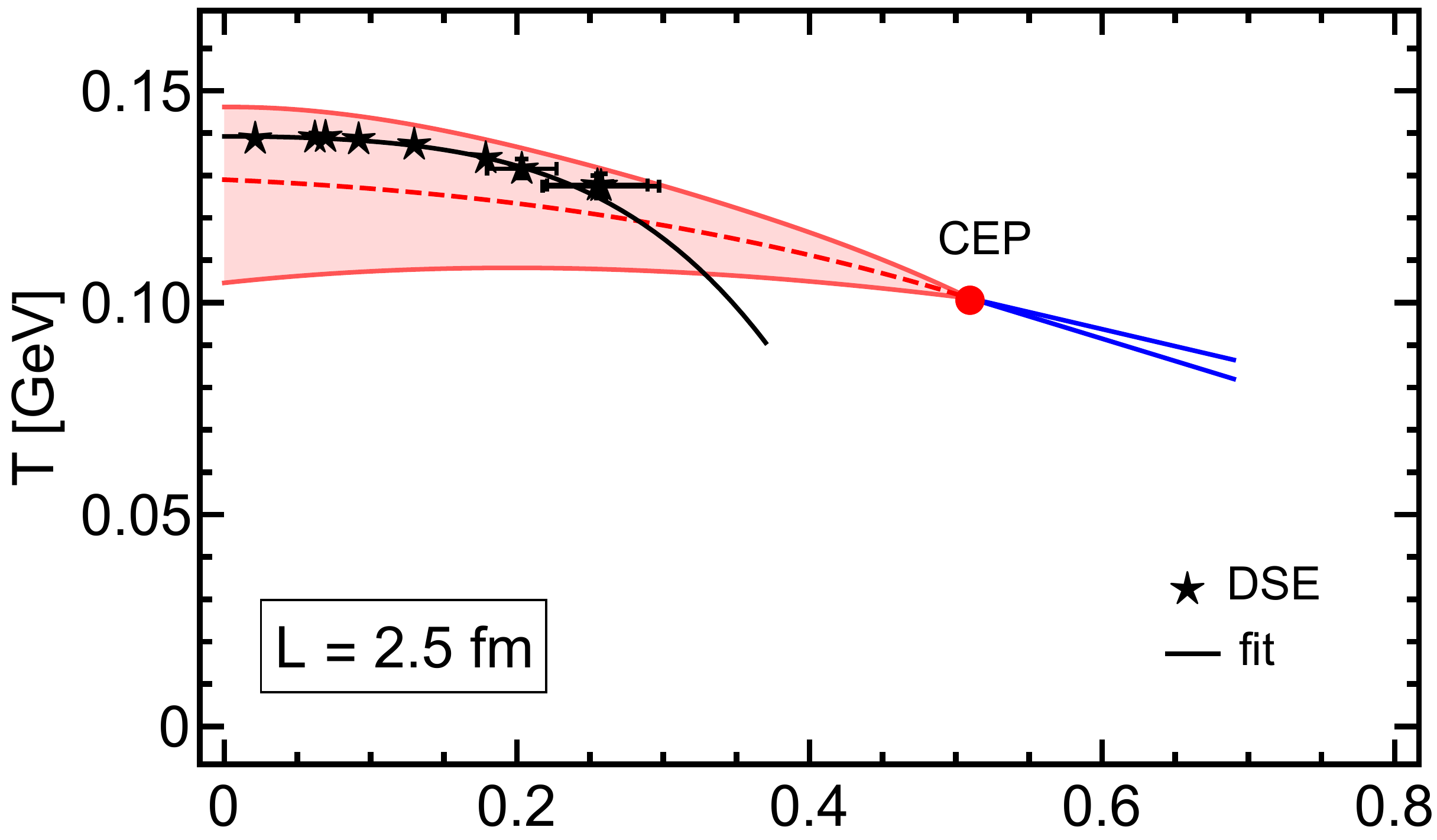}
    \includegraphics[width=0.45\textwidth]{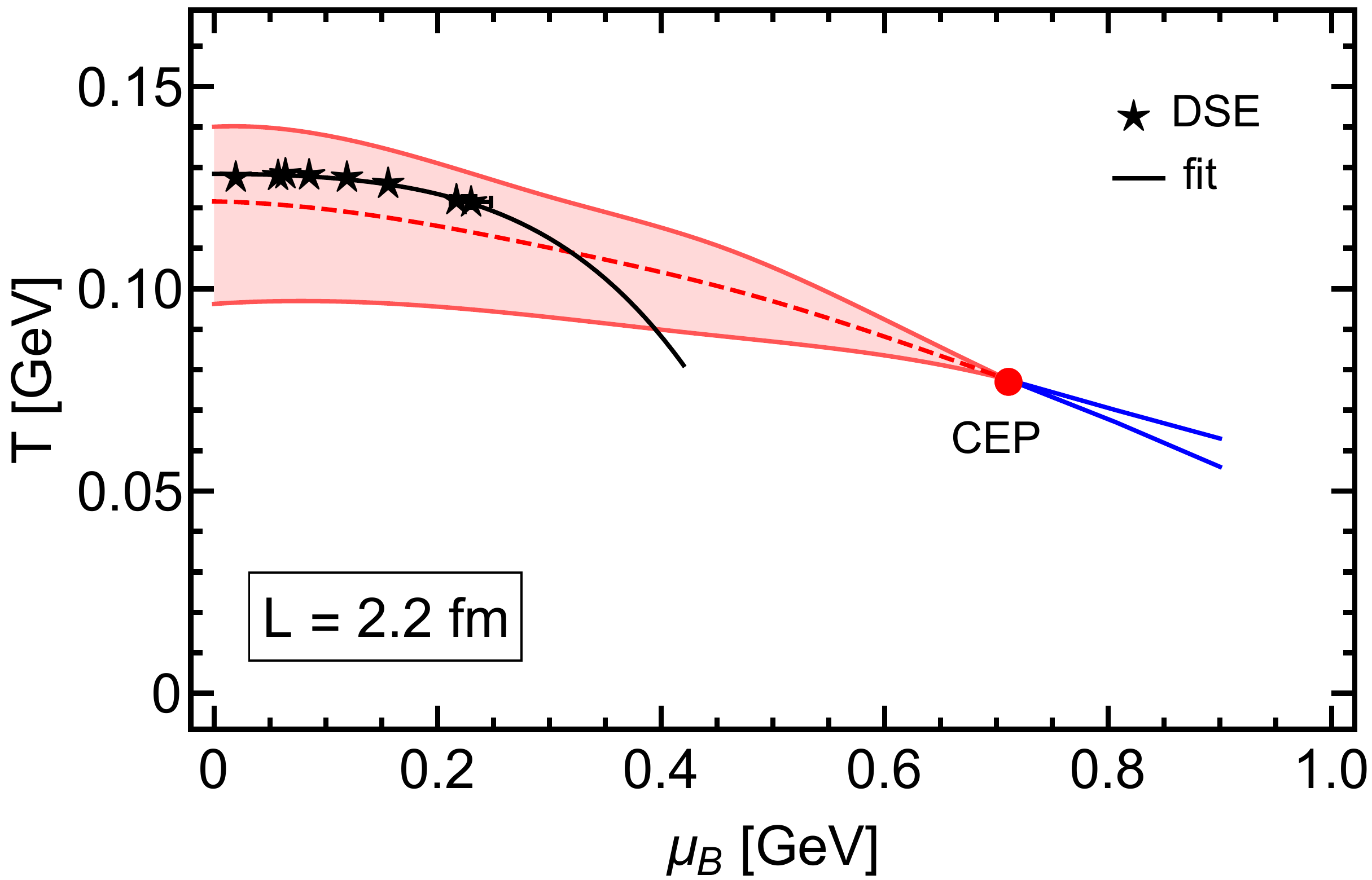}
\caption{\label{fig:phase}
(color online) Calculated QCD phase diagram in case of $L\! =\! \infty$, $2.5\,$fm and $2.2\,$fm. The red-dashed curves are the phase boundaries defined with the maximum of chiral susceptibility, the red-colored areas are the obtained crossover regions, and the black-soild curves are the fitted freeze-out lines.% (in case of $L = \infty$, we show also those given in lattice QCD~\cite{Borsanyi:2014ewa} and HRG model~\cite{Alba:2014eba} for comparison).
}
\end{figure}

%%%%% Results
\section{Phase Diagram and Further Prediction}\label{sec:PhaseDiagram}

With the quark propagator obtained by solving the DSE, we can get the temperature and chemical potential dependence of the quark condensate and the quark dynamical mass, which are commonly regarded as appropriate order parameters of chiral phase transition.
%
%Taking the chiral susceptibility criterion~\cite{Qin:2010nq} we determine the lower boundary of the chiral phase crossover region. The obtained pseudo-critical temperatures at %zero chemical potential in the two cases are listed in Table~\ref{tab:freezeout}.
%As for the upper boundary, we assign it as the set of the states for the dynamical quark mass at zero momentum to decrease to the $10\%$ of that at $T\! = \!0$ and $\mu_{B}^{} %\! = \! 0$.
%
Taking the chiral susceptibility criterion~\cite{Qin:2010nq,Gao:2016susc} we determine the upper and lower boundary of the chiral phase crossover region by the full width at half maxima (FWHM) of the susceptibility. The chiral susceptibility is defined as:
\begin{equation}
     \chi_{T}^{} = -\partial M(\tilde{\omega}_0^2,0) / \partial T \, ,
\end{equation}
where the constituent quark mass is defined as the real part of $M(\tilde{\omega}_0^2,0)$ in Eq.~(\ref{eq:MassFunction}), the same as in Sec.~\ref{sec:FiniteSize}.

The obtained crossover regions in cases of $L\! =\! \infty$, $2.5\,$fm and $2.2\,$fm are shown as the shadowed regions in Fig.~\ref{fig:phase}.
With the chiral susceptibility criterion~\cite{Qin:2010nq,Gao:2016susc} or the fluctuation criterion~\cite{Xin:2014ela},
we determine the boundaries of the first order transition region and the location of the CEP.
The obtained results in the two cases are displayed in Fig.~\ref{fig:phase}.
We illustrate also the presently obtained chemical freeze-out lines in these cases in Fig.~\ref{fig:phase}. The figure manifests that the chemical freeze-out happens in the obtained chiral crossover region. Quantitatively, the freeze-out temperature $T^{f} = 155.7\,$MeV is higher than the $T_{c}=144.0\,$MeV in the thermodynamical limit ({\it i.e.}, with $L=\infty$) at $\mu_{B}^{}=0$,
%
%but $T^{f} (\mu_{B}^{} =0) \ngtr T_{c}(\mu_{B}^{} = 0)$ in case of $L=2.2\,$fm.
%
but the deviation is smaller in case of $L=2.2\,$fm where $T^{f} = 128.4\,$MeV, $T_{c}=121.6\,$MeV at $\mu_{B}^{}=0$.
We also notice that the finite size effect shifts the location of the CEP to higher baryon chemical potential and lower temperature drastically: in case of $L= \infty$, $3.5\,$fm, $2.8\,$fm, $2.5\,$fm and $2.2\,$fm, $(\mu_{B}^\textrm{CEP},T^\textrm{CEP})$ = (296,132), (360,121), (426,111), (510,101), (711,77.5)$\,$MeV, respectively.
%, as listed in the last row of Table~\ref{tab:frzpara}.
This is consistent with the behavior given in phenomenological model calculations~\cite{Palhares:2009tf,Bhattacharyya:2012rp}.
Compared with the location of CEP calculated from functional renormalization group (FRG) approach~\cite{Gao:2020frgdse,Fu:2020frg},
it indicates that the inclusion of the finite size effect is reasonable for $2.2\,\textrm{fm} \leq L\leq2.8\,\textrm{fm}$.

\begin{figure}[!htb]
  \includegraphics[width=0.45\textwidth]{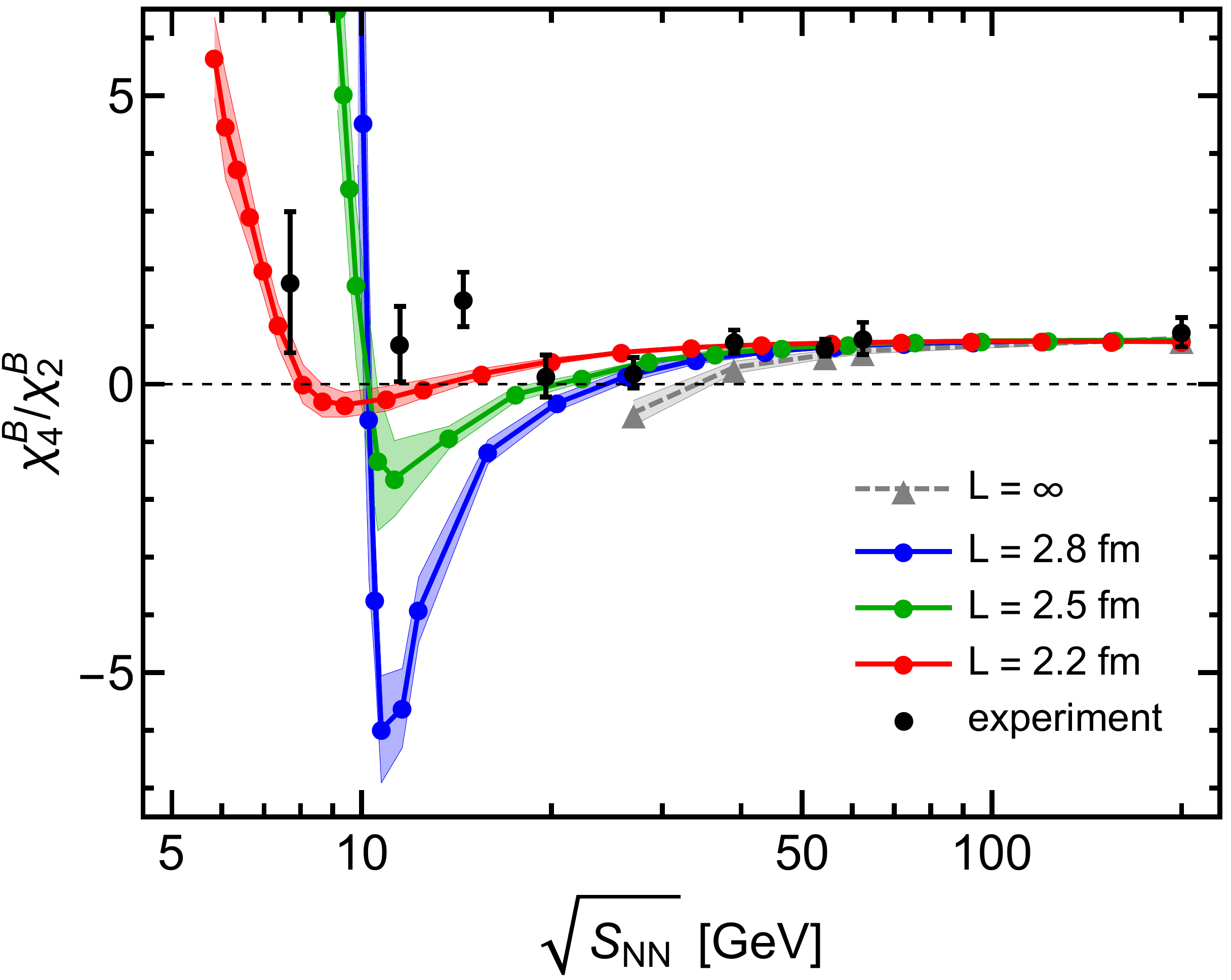}
\caption{\label{fig:x4x2}
(color online) Calculated collision energy $\sqrt{S_{NN}}$ dependence of $\kappa \sigma^2 = {\chi_{4}^{B}}/{\chi_{2}^{B}}$ at the freeze-out line.  The black circles are the experimental values~\cite{STAR:2021kur}, the gray triangles stand for our results in case of infinite volume, the blue, green and red points denote our results in the case of $L=2.8\,$fm, $2.5\,$fm and $2.2\,$fm respectively. The shadowed region(s) displays the numerical uncertainties.
}
\end{figure}

It is known that the $\kappa \sigma^2 = {\chi_{4}^{}}/{\chi_{2}^{}}$ is a direct observable in experiment and may demonstrate the property of the states around the CEP well.
We calculate ${\chi_{4}^{B}}/{\chi_{2}^{B}}$ in the $T$--$\mu_{B}^{}$ plane and pick out the value along the freeze-out line to get the $\sqrt{S_{NN}^{}}$ dependence of ${\chi_{4}^{B}}/{\chi_{2}^{B}}$.
The obtained results in case of thermodynamical limit, finite size of $L = 2.8\;$fm, $2.5\;$fm
and $2.2\,$fm are depicted in Fig.~\ref{fig:x4x2}.
In case of $L=2.2\,$fm, the lowest collision energy calculated is $5.8\,$GeV. % close to recent BES target at $3.3\,$GeV.
It is apparent that,
without considering the finite size effect, our calculated ${\chi_{4}^{B}}/{\chi_{2}^{B}}$ decreases more rapidly than the experimental data as the $\sqrt{S_{NN}^{}}$ descends.
With the finite size effect being taken into account, we can reproduce the experimental data excellently. % and the MRE correction improves the agreement a little further.
In the lower collision energy region, the $\kappa \sigma^2$ exhibits a nonmonotonic behavior, whose minimum is reached at $\sqrt{S_{NN}^{}} \approx 10\,\textrm{GeV}$, and then increases drastically as $\sqrt{S_{NN}^{}}$ further decreases.

\begin{figure}[htb]
  \includegraphics[width=0.45\textwidth]{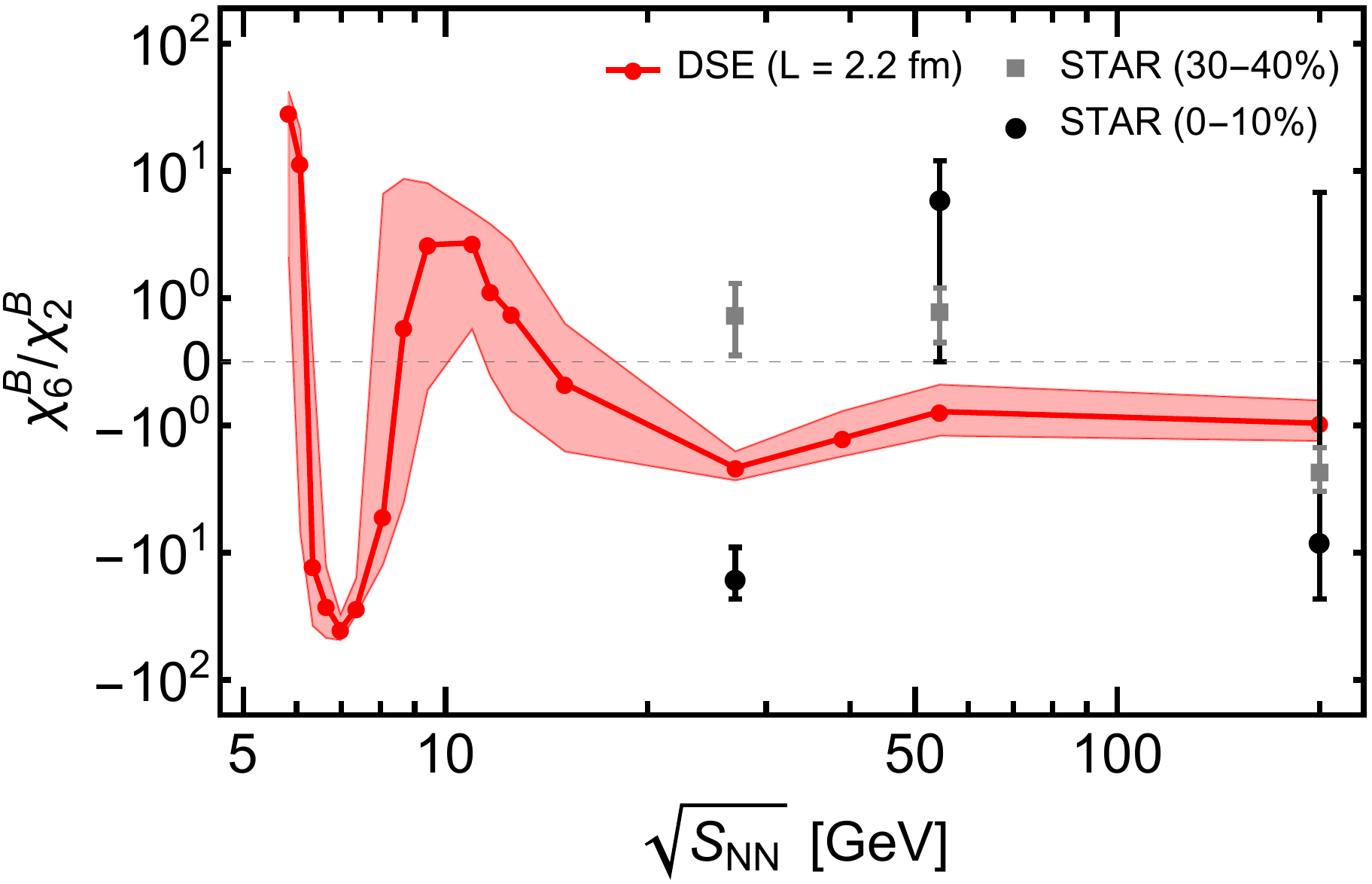}
\caption{\label{fig:x6x2}
(color online) Calculated collision energy $\sqrt{S_{NN}}$ dependence of ${\chi_{6}^{B}}/{\chi_{2}^{B}}$ (in logarithmic scale) at the freeze-out line.  
The black circles and gray squares are the experimental values~\cite{Abdallah:2021c6c2} for $0\,$-$\,10\%$ and $30\,$-$\,40\%$ centralities respectively, and the red points denote our results in case of $2.2\,$fm. The shadowed region(s) displays the numerical uncertainties.
}
\end{figure}

Since the calculated kurtosis fit the experimental data best in case of $L = 2.2\,$fm, we further calculate the $\sqrt{S_{NN}^{}}$ dependence of ${\chi_{6}^{B}}/{\chi_{2}^{B}}$ from our fitted freeze-out line under this finite size parameter, depicted in Fig.~\ref{fig:x6x2}, whose numerical uncertainties are much larger than the results of $\kappa \sigma^2$ though.

At $\sqrt{S_{NN}} = 200\,\textrm{GeV}$, we obtain a negative value of ${\chi_{6}^{B}}/{\chi_{2}^{B}} = -0.97\pm0.36$, which is qualitatively consistent with the experimental results~\cite{Pandav:2021,Nonaka:2021,Abdallah:2021c6c2} and lattice calculation~\cite{Bazavov:2020c6}.
For $\sqrt{S_{NN}} \gtrsim 10\,\textrm{GeV}$, our result shows a nonmonotonic behavior with a shallow minimum, in agreement with the FRG calculation~\cite{fu2021hyperorder}.
We also show that ${\chi_{6}^{B}}/{\chi_{2}^{B}}$ may have a large maximum along with a sharp and deep minimum when $\sqrt{S_{NN}} \lesssim 10\,\textrm{GeV}$.
As a result, we predict a complex nonmonotonic behavior of ${\chi_{6}^{B}}/{\chi_{2}^{B}}$ as a function of collision energy.

%%%%% Summary
\section{Summary}\label{sec:Summary}

In summary, we have calculated in this work the baryon number susceptibilities in a two-flavor quark system via the DSE approach of QCD in case of not only thermodynamic limit but also finite size.
By comparing the calculated ratios ${\chi_{1}^{B}}/{\chi_{2}^{B}}$ and ${\chi_{3}^{B}}/{\chi_{1}^{B}}$
with the experimental data of the net-proton multiplicity distribution in BES at RHIC, we obtained the temperature and the baryon chemical potential at the chemical freeze-out states.
We calculated also the collision energy dependence of the $\kappa \sigma^{2}$
at the freeze-out line and observed an excellent agreement with experimental data when taking into account the finite size effect.
It shows that the finite size effect is significant in studying the QCD phase transitions with RHICs.
%while the surface effect offers slight correction further.
%
The obtained collision energy $\sqrt{S_{NN}^{}}$ dependence of the $\kappa \sigma^{2}$ exhibits a nonmonotonic behavior in lower collision energy region.
We also predict that the collision energy dependence of hyper-order cumulant ratios such as ${\chi_{6}^{B}}/{\chi_{2}^{B}}$ may also be nonmonotonic.

\bigskip

\begin{acknowledgments}
The work was supported by the National Natural Science Foundation of China under Grant Nos. 11175004 and 11435001,
and the National Key Basic Research Program of China under Grant No.\ 2015CB856900.
\end{acknowledgments}

%\bibliographystyle{unsrt}
%\bibliography{Ref-FreezeOut}


\end{document}